\documentclass[referee]{aa} 
\usepackage{amsmath}

\usepackage{graphicx}
\usepackage{txfonts}
\usepackage{epsfig}

\begin{document}
   \title{A super-Eddington wind scenario for the progenitors of type Ia supernovae: binary population synthesis calculations}

   \author{B. Wang \inst{1,2}
          \and
          X. Ma \inst{1,2}
          \and
          D.-D. Liu \inst{1,2}
          \and
          Z.-W. Liu \inst{3}
          \and
          C.-Y. Wu \inst{1,2}
          \and
          J.-J. Zhang \inst{1,2}
          \and
          Z. Han \inst{1,2}
          }

   \institute{Yunnan Observatories, Chinese Academy of Sciences, Kunming 650026, China;
              \email{wangbo@ynao.ac.cn}
              \and
              Key Laboratory for the Structure and Evolution of Celestial Objects, Chinese Academy of Sciences, Kunming 650026, China
              \and
              Argelander-Institut f\"{u}r Astronomie, Auf dem H\"{u}gel 71, D-53121, Bonn, Germany
              }
   \date{Accepted 23 February 2015}

\abstract
{The super-Eddington wind scenario has been proposed as an alternative way for producing type Ia supernovae (SNe
Ia). The super-Eddington wind can naturally prevent the carbon--oxygen white dwarfs (CO WDs) with high mass-accretion rates from
becoming red-giant-like stars. Furthermore, it works in low-metallicity environments, which may explain SNe Ia
observed at high redshifts.}
{In this article, we systematically investigated the most prominent single-degenerate
WD+MS channel based on the super-Eddington wind scenario. }
{We combined the Eggleton stellar evolution code with a rapid  binary population synthesis (BPS) approach to predict SN Ia birthrates for the WD+MS channel by adopting the super-Eddington wind scenario and detailed mass-accumulation efficiencies of H-shell flashes on the WDs.}
{Our BPS calculations found that the estimated SN Ia birthrates for the WD+MS channel
are $\sim$0.009$-$0.315$\times10^{-3}\,{\rm yr}^{-1}$ if we adopt the Eddington accretion rate as the critical accretion rate, which are much lower than that of the observations ($<$10\% of the observed SN Ia birthrates). This indicates that the WD+MS channel only contributes a small proportion of all SNe Ia. The birthrates in this simulation are lower than previous studies, the main reason of which is that new mass-accumulation efficiencies of H-shell flashes are adopted. We also found that the critical mass-accretion rate has a significant influence on the birthrates of SNe Ia. Meanwhile, the results of our BPS calculations are sensitive to the values of the common-envelope ejection efficiency.}
{}


\titlerunning{A super-Eddington wind scenario of SNe Ia: BPS calculations}

\authorrunning{B. Wang et al.}

   \maketitle

%

\section{Introduction} \label{1. Introduction}

Type Ia supernovae (SNe Ia) have been successfully used as standard
candles for measuring cosmological distances, resulting
in the determination of the accelerating expansion of the
Universe that is driven by dark energy (e.g., Riess et al. 1998;
Perlmutter et al. 1999). There is a theoretical consensus that
SNe Ia are thermonuclear explosions of carbon-oxygen white
dwarfs (CO WDs) (see Hoyle \& Fowler 1960; Nomoto et al. 1997).
However, SN Ia progenitor
systems and the physics of their explosion mechanism are
still uncertain (for recent reviews see Wang \& Han 2012;
Hillebrandt et al. 2013; Maoz et al. 2014; Ruiz-Lapuente 2014), which
may affect the precision of results of the current cosmological
model (e.g., Howell 2011).

Depending on the nature of mass donor stars, two ways by
which CO WDs in binary systems can accrete mass towards the
Chandrasekhar limit and then cause SNe Ia have been proposed, i.e.,
the single-degenerate (SD) model and the double-degenerate
(DD) model. In the SD model, a CO WD accretes material
from a non-degenerate star to increase its mass to the
Chandrasekhar mass limit, which then leads to an SN Ia explosion
(e.g., Whelan \& Iben 1973; Nomoto et al. 1984).
In the DD model, two CO WDs in a close binary  are brought
together by the gravitational wave radiation and merge to
result in an SN Ia explosion (e.g., Webbink 1984; Iben \& Tutukov 1984;
Nelemans et al. 2001; Geier et al. 2007; Ruiter et al. 2009; Chen et al. 2012).
For recent observational constraints to theoretical models see, e.g., Maoz et al. (2014),
Wang \& Han (2012), Liu et al. (2012a) and Wang et al. (2013b).

The most commonly-considered variant in the SD model is the
``WD+MS channel'' (usually known as the ``supersoft channel''),
in which the WD accretes material from a main-sequence (MS) star
or a slightly evolved sub-giant star  (e.g., Hachisu et al. 1996, 1999;
Li \& van den Heuvel 1997;  Langer et al.\ 2000; Han \& Podsiadlowski
2004; Wang et al. 2010; Claeys et al. 2014; Ablimit et al. 2014).
This channel has been theorized to be accompanied by recurrent novae
and supersoft X-ray sources in the past two decades (see Parthasarathy
et al.\ 2007; van den Heuvel et al. 1992; Rappaport et al. 1994). Note
that we did not study the SD WD+RG channel in this paper as the SN Ia
birthrate from that channel is far below the birthrate from the WD+MS
channel (e.g., Yungelson \& Livio 1998; L\"{u} et al. 2006; Mennekens
et al. 2010; Claeys et al. 2014).

The accretion of H-rich material on WDs
is crucial for understanding SNe Ia from the WD+MS channel.
If the mass-accretion
rate is too high, the WD will evolve into a red-giant-like star
due to the pileup of the accreted material on its surface. If
the mass-accretion rate is too low, the nuclear burning is
unstable, H-shell flashes occur and expel the accreted shell.
Thus, there is only a narrow range where the accreted H-rich
material could steadily burn on the surface of WDs.
This leads to a fact that theoretical SN Ia birthrates from the WD+MS channel are
far below those of observations (e.g., Yungelson et al. 1996).
The optically thick wind assumption was proposed to solve
this problem, in which the H-rich material is transformed into
He at a critical rate, while the unprocessed material is
blown off in the form of the optically thick wind (see
Hachisu et al. 1996, 1999; Nomoto et al. 2007). However,
the optically thick wind assumption is still quite controversial
(see Langer et al. 2000). Besides, this assumption does not work
when the metallicity is lower than 0.002 (see Kobayashi et al.
1998), while, as a matter of fact, SNe Ia with a  metallicity
lower than this threshold have been observed (e.g., Prieto et
al. 2008; Badenes et al. 2009).

Ma et al. (2013) recently proposed an alternative way
to replace the optically thick wind regimes, which is called
¡°the super-Eddington wind scenario¡±. In this scenario, the
super-Eddington wind is triggered when the luminosity of the
accreting WD exceeds the Eddington luminosity. The super-Eddington
wind can naturally prevent the CO WDs with high mass-accretion
rates from forming red-giant-like stars. In contrast to the
optically thick wind, the super-Eddington wind does not
significantly depend on the metallicity, which could
be triggered even for $Z=10^{-6}$ (see Fig. 4 of Ma et al. 2013).
This indicates that this scenario may have a contribution to SNe
Ia at high redshifts. In this work, we aim to investigate theoretical
SN Ia birthrates and delay-time distributions of the WD+MS channel
by adopting the super-Eddington wind scenario.

As mentioned above, at low mass-accretion rates, the accreting
WD would experience H-shell flashes like nova outbursts.
However, the H-shell flashes have not been well understood due
to the numerical difficulties in nova outbursts. For example,
only one cycle of H-shell flash has been followed in most
simulations of nova outbursts; multicycle evolution of the
H-shell flash is difficult to compute (e.g., Kovetz \& Prialnik
1985; Iben et al. 1992; Schwartzman et al. 1994).
The H-shell flashes are important for understanding the SN Ia
progenitor scenario studied in this paper; a low mass-accumulation
efficiency of H-shell flashes will lead to a low birthrate of SNe Ia.
Denissenkov et al. (2013a) recently constructed multi-cycle nova
evolutionary sequences with WDs using the stellar evolution code
\textit{Modules for Experiments in Stellar Astrophysics} (MESA).
Adopting a similar method to Denissenkov et al. (2013a),
Ma et al. (2013) obtained the mass-accumulation efficiency of
H-shell flashes for a wide range of WD masses.
A more detailed binary population synthesis (BPS)
approach is needed to study the influence of the H-shell
flashes on the final results and to give SN Ia birthrates, which is
also addressed in this work.

This paper is organized as follows. The numerical code for
the binary evolution calculations is described in Sect.\,2, and the
corresponding results are provided in Sect.\,3. The BPS method and
results  are presented in Sect.\,4. Our discussion and summary are then
presented in Sect.\,5.

\section{Numerical code for binary evolution}

In the WD+MS channel, the mass donor star is a MS star or
a slightly evolved sub-giant star, which transfers H-rich material
on the surface of the WD. The accreted H-rich material
is burned into He, and then the He is converted into carbon and
oxygen, resulting in the mass increase of the WD.
We assume that the WD explodes as an SN Ia once its mass grows to
1.378$\,M_{\odot}$, which is the critical mass limit of non-rotating
WDs for carbon ignition (e.g., Nomoto et al. 1984).\footnote{We did
not consider the influence of rotation on the evolution of the
accreting WD in this work; the maximum stable mass for a rotating WD
may be above the standard Chandrasekhar mass limit (see Uenishi et al.
2003; Yoon et al. 2004; Wang et al. 2014). }
We calculated the evolution of the WD+MS systems using the Eggleton
stellar evolution code (Eggleton 1973;  Han et al. 1994; Pols et al.
1995, 1998; Eggleton \& Kiseleva-Eggleton 2002). The initial setup
and basic input physics for this code are similar to those in Wang
et al. (2010). The MS star models in our calculations are composed
of metallicity $Z=0.02$, H abundance $X=0.70$, and He abundance $Y=0.28$.

\begin{figure}
\begin{center}
\epsfig{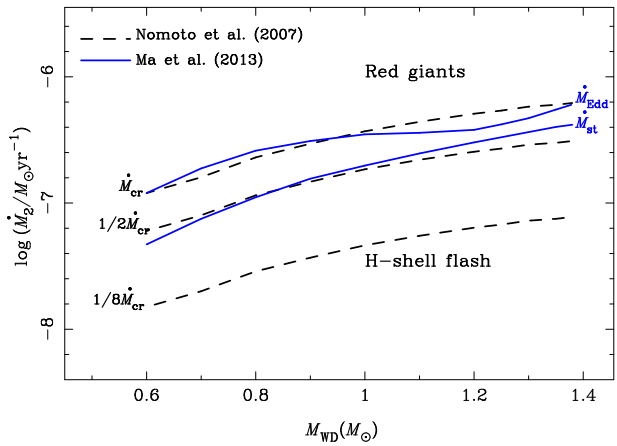}
\caption{Properties of H-shell burning on the surface of accreting WDs in the
plane of WD mass and mass-accretion rate. A comparison of the critical mass-accretion rate between
the super-Eddington wind scenario and the optically thick wind model was made.
The solid lines are from Ma et al. (2013), while the
dashed lines are from Nomoto et al. (2007). }
\end{center}
\end{figure}

The mass transfer begins once the MS star fills its Roche-lobe. We define
the values of mass-accumulation efficiency for H-shell burning on the
surface of the WD, $\eta _{\rm H}$, as
 \begin{equation}
\eta_{\rm H}=\left\{
 \begin{array}{ll}
 \dot{M}_{\rm Edd}/|\dot{M}_{\rm 2}|, & |\dot{M}_{\rm 2}|> \dot{M}_{\rm
Edd},\\
 1, & \dot{M}_{\rm Edd}\geq |\dot{M}_{\rm 2}|\geq\dot{M}_{\rm st},\\
\eta'_{\rm H}, & |\dot{M}_{\rm 2}|< \dot{M}_{\rm st},
\end{array}\right.
\end{equation}
where $\dot{M}_{\rm 2}$, $\dot{M}_{\rm Edd}$ and $\dot{M}_{\rm st}$ are
the mass-transfer rate, the Eddington critical accretion rate,  and the
minimum accretion rate for stable H-shell burning, respectively.
In the following, we make a detailed description for the above equation:
(1) If the mass-transfer rate, $\dot{M}_{\rm 2}$,
is above the Eddington critical accretion rate, $\dot{M}_{\rm Edd}$, we
assume that the accreted H steadily burns on the surface of the WD, and
that the H-rich material is converted into He at a rate of
$\dot{M}_{\rm Edd}$,
 \begin{equation}
\dot{M}_{\rm Edd}=5.975\times10^{-6}({M}_{\rm WD}^4-3.496{M}_{\rm WD}^3+4.373{M}_{\rm WD}^2-2.226{M}_{\rm WD}+0.406),
 \end{equation}
where ${M}_{\rm WD}$ is the mass of the WD in units of $M_{\odot}$, and
$\dot{M}_{\rm Edd}$ is in units of  $M_{\odot}{\rm yr}^{-1}$.
The unprocessed material can be lost from the system in the form of the super-Eddington wind
with a mass-loss rate of $(|\dot{M}_{\rm 2}|-\dot{M}_{\rm Edd})$. In Fig.\,1, we make
a comparison of the critical mass-accretion rate between
the super-Eddington wind scenario and the optically thick wind model (for
more discussions about this figure see Sect.\,5).
(2) When $|\dot{M}_{\rm 2}|$ is smaller than $\dot{M}_{\rm Edd}$
but higher than the minimum accretion rate for stable H-shell burning,
$\dot{M}_{\rm st}$, the burning of the H layer  is steady and no mass
is lost from the system. The values of $\dot{M}_{\rm st}$ are
 \begin{equation}
\dot{M}_{\rm st}=3.057\times 10^{-7}({M}_{\rm WD}^2-0.386{M}_{\rm WD}+0.027),
 \end{equation}
where $\dot{M}_{\rm st}$ is in units of  $M_{\odot}{\rm yr}^{-1}$ (see Fig.\,1).
(3) H-shell flashes are triggered once $|\dot{M}_{\rm 2}|$ is lower than $\dot{M}_{\rm st}$.
The mass-accumulation efficiency for H-shell flashes, $\eta'_{\rm H}$, is
linearly interpolated from a detailed grid (see Fig. 2), in which a wide range
of WD masses and mass-accretion rate were calculated by Ma et al. (2013).

\begin{figure}
\begin{center}
\includegraphics[width=8.9cm,angle=0]{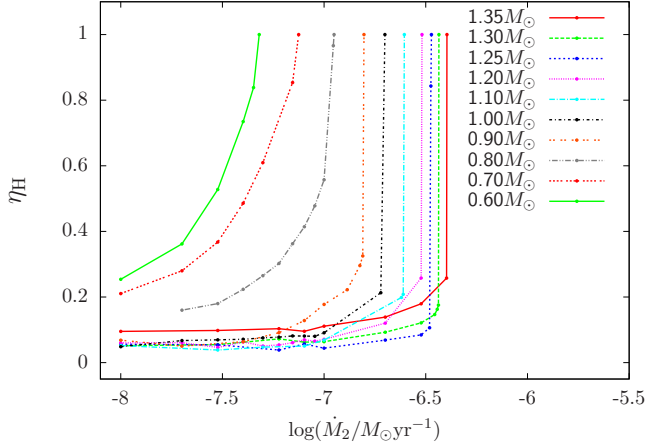}
\caption{The mass-accumulation efficiency of H-shell flashes, $\eta _{\rm H}$, is plotted against the
mass-accretion rate for various initial WD masses. The data points are taken from Ma et al. (2013). }
  \end{center}
\end{figure}

Helium is assumed to be ignited once the mass of the He layer under the H-shell
burning reaches a certain value. The mass-growth rate of the He layer is defined as
 \begin{equation}
 \dot{M}_{\rm He}=\eta _{\rm H}|\dot{M}_{\rm 2}|.
 \end{equation}
If He-shell
flashes occur, we assume that a part of the layer mass is blown off. In
this case, we adopt the prescription of Kato \& Hachisu (2004)
for mass-accumulation efficiency of He-shell flashes.\footnote{Yoon et al.
(2004) argued that the burning of He layer is less violent when rotation is
considered, which may increase the mass-accumulation efficiency of He-rich
material.} We define the mass-growth rate of the WD, $\dot{M}_{\rm CO}$, as
 \begin{equation}
 \dot{M}_{\rm CO}=\eta_{\rm He}\dot{M}_{\rm He}=\eta_{\rm He}\eta_{\rm
 H}|\dot{M}_{\rm 2}|,
  \end{equation}
where $\eta_{\rm He}$ is the mass-accumulation efficiency of He-shell flashes.

The evolution of the orbital periods of these WD+MS systems
is mainly decided by the wind mass-loss of the WD, which
carries away the orbital angular momentum of the binaries.
We assume that the mass lost from these systems takes away
the specific orbital angular momentum of the WD, but the mass
lost from the wind of the mass donors
is negligible (e.g., Wang et al. 2010). Finally,
dense model grids of binary evolution calculations are obtained.

\section{Binary evolution results}

\begin{figure*}
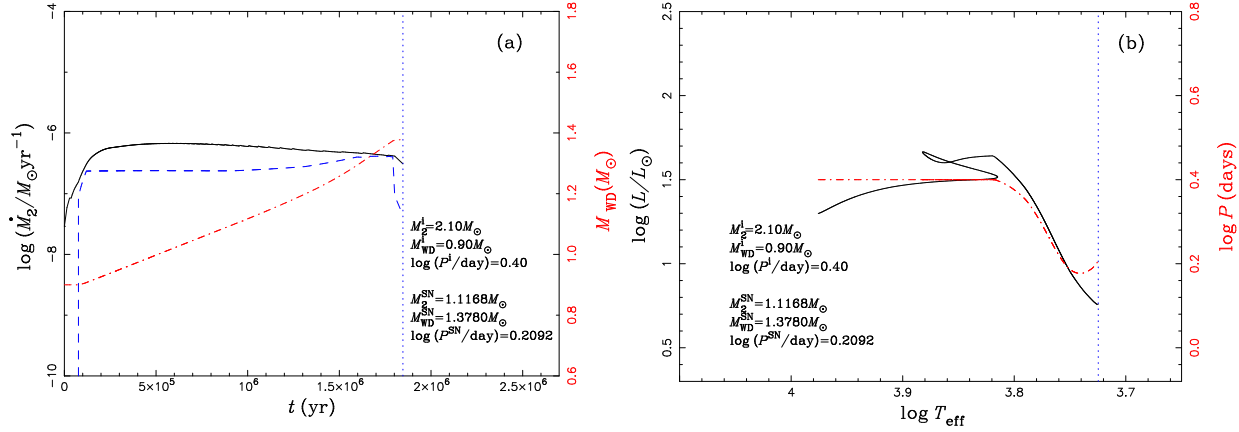

\centerline{\epsfig{file=f3a.ps,angle=270,width=8cm}\ \
\epsfig{file=f3b.ps,angle=270,width=8cm}} \caption{A representative
example of binary evolution calculations, where the binary system is
in the weak H-shell flash phase at the moment of the SN explosion.
In panel (a), the solid, dashed and dash-dotted curves present
$\dot M_2$, $\dot M_{\rm CO}$ and $M_{\rm WD}$,
varying with time, respectively. In panel (b), the evolutionary track
of the donor in the Hertzsprung-Russell diagram is shown as a solid
curve, and the evolution of orbital period is shown as a dash-dotted
curve. Dotted vertical lines in both
panels indicate the position where the WD produces an SN Ia.}
\end{figure*}

In Fig.\,3, we present a representative example of binary
evolution calculations. The binary system is in the weak
H-shell flash phase at the moment of SN explosion.
The initial condition for this binary is ($M_2^{\rm i}$, $M_{\rm WD}^{\rm i}$, $\log
(P^{\rm i}/{\rm day})$) $=$ (2.1, 0.9, 0.4), where $M_2^{\rm i}$,
$M_{\rm WD}^{\rm i}$ and $P^{\rm i}$ are the initial masses of the MS
star and the CO WD in solar masses, and the initial orbital
period in days, respectively. The MS star firstly fills its
Roche-lobe after the exhaustion of central H (it now contains a He
core), which leads to early Case B mass transfer; the orbital period of
the binary system is not changed until the MS star fills its Roche-lobe,
see panel (b) of Fig.\,3. The mass-transfer rate $|\dot{M}_{\rm 2}|$ exceeds $\dot
M_{\rm Edd}$ soon after the onset of Roche-lobe overflow, resulting in a wind
phase in which a part of the transferred mass is blown off in the form of the
super-Eddington wind, and the remainder is accumulated on the WD; the wind
carries away the specific orbital angular momentum of the WD, which results
in the decrease of the orbital period. After about
$1.6\times10^{6}$\,yrs, $|\dot{M}_{\rm 2}|$ drops below $\dot M_{\rm
Edd}$, but is still higher than $\dot M_{\rm st}$. At this stage, the H-shell burning is stable, but the
super-Eddington wind stops. With the
decrease of $|\dot{M}_{\rm 2}|$, the WD quickly enters
a weak H-shell flash phase. The orbital period of the binary starts to increase
once the accreting WD is more massive than the donor star. The WD
always grows in mass until it explodes as an SN Ia in the weak
H-shell flash phase. When the WD increases its mass to $1.378\,M_{\odot}$, the mass of the
donor star is $M^{\rm SN}_2=1.1168\,M_{\odot}$ and the orbital
period is $\log (P^{\rm SN}/{\rm day})=0.2092$.

\begin{figure}
\begin{center}
\epsfig{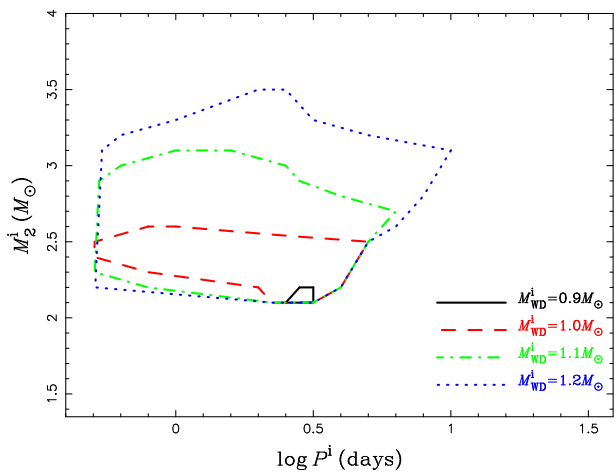} \caption{Regions in the $\log P^{\rm i}-M^{\rm i}_2$
plane for WD+MS binaries that ultimately produce SNe Ia for various initial
WD masses, where $P^{\rm i}$ and $M^{\rm i}_2$ are the initial orbital period and the initial
mass of the MS star, respectively. }
\end{center}
\end{figure}

Figure 4 presents the initial contours for producing SNe Ia in
the orbital period and secondary mass plane for various initial WD masses (i.e.,
$M_{\rm WD}^{\rm i}$ = 0.90, 1.0, 1.1 and $1.2\,M_{\odot}$).
The enclosed region almost vanishes for WDs with
$0.9\,M_{\odot}$, which is then assumed to be the minimum initial WD
mass for producing SNe Ia from the WD+MS channel.
We note that the critical mass-accretion rate for WDs is still uncertain (e.g., Cassisi et al. 1998;
Langer et al.\ 2000; Nomoto et al. 2007; Shen \& Bildsten 2007; Bours et al. 2013).
In the observations, some binaries which contain a neutron star or a black hole can lose their masses
at rates that exceed  the Eddington accretion rate of
the compact object by several times (e.g., Cyg X-2 and SS 433;
Podsiadlowski \& Rappaport 2000; Blundell et al. 2001). In Fig.\,5,
we show the initial contours for producing SNe Ia
after adopting $2\dot{M}_{\rm Edd}$ as the critical mass-accretion rate;
we did not change the values of the minimum accretion rate for stable
H-shell burning (i.e., $\dot{M}_{\rm st}$) in our calculations.
This is an unphysical extreme assumption, but we can use this to examine the influence of the
critical mass-accretion rate on the final results. From this figure,
we can see that the initial contours for producing SNe Ia are enlarged significantly. This is because
a larger regime for stable H-shell burning is adopted.

\begin{figure}
\begin{center}
\epsfig{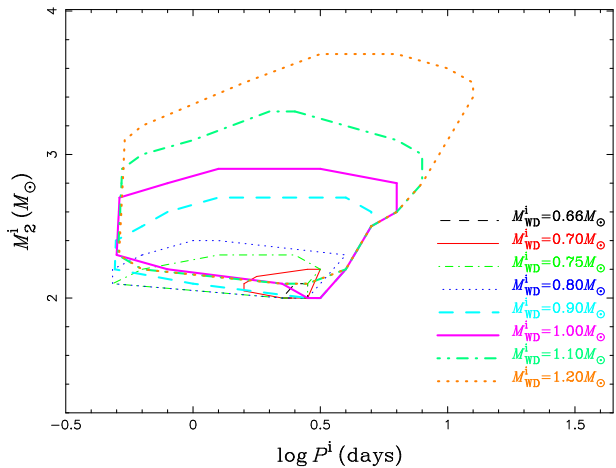} \caption{Similar to Fig. 4, but for
results obtained after adopting
$2\dot{M}_{\rm Edd}$ as the critical mass-accretion rate.}
\end{center}
\end{figure}

The left boundaries of the initial contours in Figs\,4--5 are constrained by
the condition that Roche-lobe overflow
begins when the mass donor star is in the zero-age MS stage, while systems beyond the
right boundary experience mass transfer at a relatively high rate due to
the rapid expansion of the donor stars in the Hertzsprung gap
and they lose too much mass
via the super-Eddington wind, preventing the WDs from increasing their
masses to the Chandrasekhar mass. The upper boundaries are set by a high
mass-transfer rate owing to a large mass ratio between the mass donor star
and the WD, while the lower boundaries
are constrained by the condition that the mass-transfer rate should be high
enough to ensure the mass growth of the WD during H-shell flashes.

\section{BPS assumptions and results}

\subsection{Method of BPS}
In order to obtain SN Ia birthrates and delay times for the WD+MS channel,
a series of Monte Carlo BPS simulations are performed.
The random number generator algorithm in our Monte Carlo simulations
is from Press et al. (1992), which provides a uniform probability density in the range of (0, 1).
The repetition period of the random number generator in our simulations is $\gtrsim$$10^{18}$,
which is infinite for practical calculations.

For each BPS simulation, we employed the
Hurley rapid binary evolution code (Hurley et al. 2002) to simulate
the evolution of $10^{\rm 7}$ primordial binaries,
and adopted a metallicity of $Z=0.02$. These binaries are followed from
the star formation to the formation of the WD+MS systems based on
three binary evolutionary ways (see Fig. 1 of Wang \& Han 2012).
If the initial parameters of a CO WD+MS system at the onset of the
Roche-lobe overflow are
located in the SN Ia production regions in the plane of
($\log P^{\rm i}$, $M_2^{\rm i}$) for its specific $M_{\rm WD}^{\rm i}$,  then an
SN Ia is assumed to occur. We adopt linear interpolation if
$M_{\rm WD}^{\rm i}$ is not among the masses listed in Figs 4--5.

Seven sets of Monte Carlo BPS simulations are conducted to investigate
the birthrates of SNe Ia (see Table 1), in which sets 1 and 2
are two best models from Camacho et al. (2014) that result in a good overall
fit to the properties of the WD+MS binary population in the Sloan Digital
Sky Survey (SDSS) Data Release 7. The following basic assumptions for
these Monte Carlo simulations are adopted:

(1) A constant star formation rate (SFR) of  $5\,{M}_{\odot}\rm yr^{-1}$
is simply assumed over the past 14 Gyrs or, alternatively, it is
modeled as a delta function, that is a single instantaneous starburst
($10^{10}\,M_{\odot}$ in stars is assumed). We intend the constant SFR to approximate spiral galaxies,
and the delta function to provide a rough description of elliptical
galaxies or globular clusters (e.g., Wang et al. 2013a).

(2) All stars are assumed to be members of binaries.
The primordial binary samples are generated via a Monte Carlo method and
an initially circular orbit was supposed for all binaries. The distribution
of initial orbital separations is assumed to be constant in $\log a$ for
wide binaries, where $a$ is the orbital separation (e.g., Han et al. 1995).

(3) The initial mass function (IMF) of the primordial primary star is taken
from Miller \& Scalo (1979, MS79). Alternatively, we also consider the IMF
of Scalo (1986, S86).

(4) A constant mass-ratio distribution is taken, i.e., $n(q)=1$, or a
distribution proportional to the mass ratio, $n(q)\propto q$ (e.g.,
Camacho et al. 2014).

(5) The common-envelope (CE) evolution is still uncertain (e.g., Zuo \& Li 2014).
The standard energy equations are used to describe the output during
the CE stage (e.g., Webbink 1984).
There are two uncertain
parameters for this prescription of the CE ejection, i.e.,
$\alpha_{\rm CE}$ (the CE ejection efficiency) and $\lambda$
(a structure parameter that depends on the evolutionary stage of the mass
donor). $\lambda$ is usually set to be 0.5 for the purpose of constraining
$\alpha_{\rm CE}$ (see de Kool 1990). Here, we set $\alpha_{\rm CE}$ to be
0.3, 1.0 and 3.0 to examine its influence on the final results.

\subsection{Results of BPS}

\begin{table}
\begin{center}
 \begin{minipage}{85mm}
 \caption{Galactic SN Ia birthrates for seven simulation sets, in which
sets 1 and 2 are two best models from Camacho et al. (2014). Notes:
$\alpha_{\rm CE}$ = CE ejection efficiency; $n(q)$ = initial mass-ratio
distribution; ${\rm IMF}$ = initial mass function; $\dot{M}_{\rm cr}$ =
critical mass-accretion rate; $\nu$ = Galactic SN Ia birthrate.}
\begin{tabular}{ccccccc}
\hline \hline
Set & $\alpha_{\rm CE}$ & $n(q)$ & ${\rm IMF}$ & $\dot{M}_{\rm cr}$ & $\nu$ ($10^{-3}$\,yr$^{-1}$)\\
\hline
$1$ & $0.3$     & ${\rm 1}$      & ${\rm MS79}$  & $\dot{M}_{\rm Edd}$  & $0.315$\\
$2$ & $0.3$     & $\propto{q}$   & ${\rm MS79}$  & $\dot{M}_{\rm Edd}$  & $0.137$\\
$3$ & $1.0$     & ${\rm 1}$      & ${\rm MS79}$  & $\dot{M}_{\rm Edd}$  & $0.110$\\
$4$ & $1.0$     & ${\rm 1}$      & ${\rm S86}$   & $\dot{M}_{\rm Edd}$  & $0.070$\\
$5$ & $1.0$     & ${\rm 1}$      & ${\rm MS79}$  & $2\dot{M}_{\rm Edd}$ & $0.955$\\
$6$ & $3.0$     & ${\rm 1}$      & ${\rm MS79}$  & $\dot{M}_{\rm Edd}$  & $0.009$\\
$7$ & $3.0$     & ${\rm 1}$      & ${\rm MS79}$  & $2\dot{M}_{\rm Edd}$ & $0.712$\\
\hline
\end{tabular}
\end{minipage}
\end{center}
\end{table}

\begin{figure}
\begin{center}
\epsfig{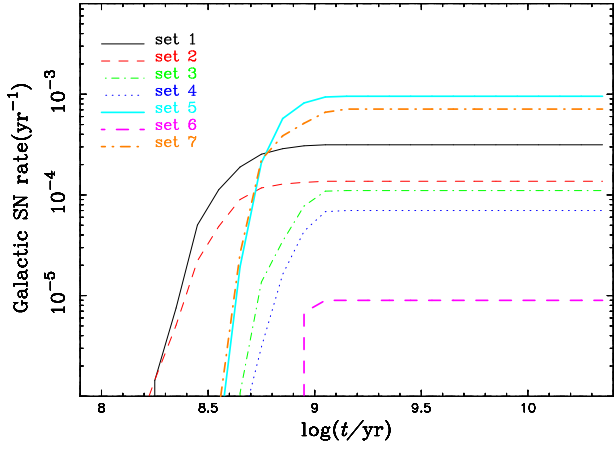}
\caption{Evolution of SN Ia birthrates for a constant star-formation rate
(${\rm SFR}=5.0\,M_{\rm \odot}{\rm yr}^{-1}$) with different BPS simulation sets.}
\end{center}
\end{figure}

Based on the seven sets of simulations for the WD+MS channel,
the estimated birthrates of SNe Ia are strongly sensitive to the choice of
some initial conditions, e.g., the critical mass-accretion rate,
the CE ejection efficiency, initial mass-ratio distribution and
initial mass function, etc. Significantly,
if we adopt $2\dot{M}_{\rm Edd}$ as the critical mass-accretion rate,
the SN Ia birthrate will increase to be a maximum value. This is because a larger
initial contours for producing SNe Ia are adopted.

In Fig.\,6, we show the evolution of SN Ia birthrates with time by adopting
metallicity $Z=0.02$ and star-formation rate ${\rm SFR}=5\,M_{\rm \odot}{\rm yr}^{-1}$.
This study presents an SN Ia birthrate of $\sim$$0.009-0.315\times 10^{-3}\,{\rm
yr}^{-1}$ by adopting $\dot{M}_{\rm Edd}$ as the critical accretion rate, which
is far below the observed value
(i.e., $3-4\times 10^{-3}\ {\rm yr}^{-1}$; Cappellaro \& Turatto 1997).\footnote{The
upper mass limit for CO WDs has been suggested to be $\sim$$1.07\,M_\odot$ (see
Umeda et al. 1999). If this mass limit is adopted in Fig.\,4, the SN Ia birthrate
from the WD+MS channel will decrease to $\sim$$0.009-0.215\times10^{-3}\,{\rm
yr}^{-1}$.}
Thus, we speculate that the WD+MS channel only contributes to a small part
of all SNe Ia ($<$10\%); some other formation channels or mechanisms may also have a
contribution to SNe Ia (for a recent review see Wang \& Han 2012).
If we adopt $2\dot{M}_{\rm Edd}$ as the critical accretion rate,
the birthrate of SNe Ia will increase to $\sim$$0.712-0.955\times
10^{-3}\,{\rm yr}^{-1}$ (see sets 5 and 7). This
value is still lower than that of observations, but it provides an upper limit on
the birthrate of SNe Ia for the WD+MS channel. This means that the variation of the critical
accretion rate has a significant influence on the SN Ia birthrates.

\begin{figure}
\begin{center}
\epsfig{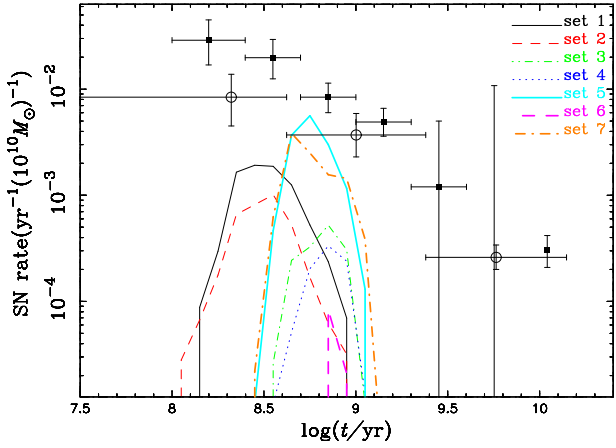}
\caption{Delay-time distributions of SNe Ia with different BPS simulation sets.
The open circles are taken from Maoz et al. (2011), while the filled squares are from Totani et al. (2008).}
\end{center}
\end{figure}

\begin{figure}
\begin{center}
\epsfig{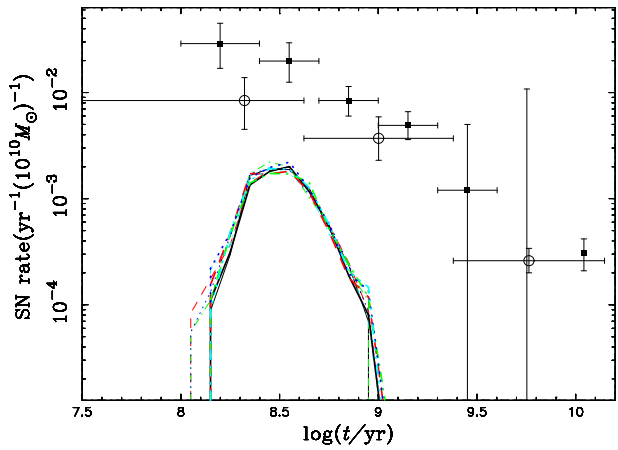}
\caption{Similar to Fig.\,7, but for the case of set 1,
where ten random initial seeds are adopted in our simulations.  }
\end{center}
\end{figure}

SN Ia delay times are defined as the time interval from star formation to the SN explosion.
In Fig.\,7,  we display the SN Ia delay-time distributions for the WD+MS channel
with different BPS simulation sets, which are obtained from a single starburst.
From this figure, we can see that the
estimated delay times for this channel are $\sim$110\,Myrs$-$1200\,Myrs after
the starburst, which may have a contribution to the SNe Ia with intermediate delay times.
Note that a low value of $\alpha_{\rm CE}$ results in a higher birthrate of SNe Ia.
This is because, for a low value of $\alpha_{\rm CE}$, the primordial binary systems
need to release more orbital energy to eject the CE, and then produce WD+MS binaries.
In this case, WD+MS binaries tend to have slightly closer orbital separations, and are more likely
to be located in the SN Ia production region (e.g., Figs\,4--5). We note that,
in any Monte  Carlo  simulations the  initial  seed  of the  random  number
generator results in a given and specific simulated population.
In Fig.\,8, we show the delay-time distributions of SNe Ia but for the case of set 1, in which
ten random initial seeds are adopted in our simulations. From this figure, we found that
there is no significant difference among these simulations, i.e., our Monte  Carlo BPS results are not
sensitive to the initial seed of the random number generator.

\begin{figure*}
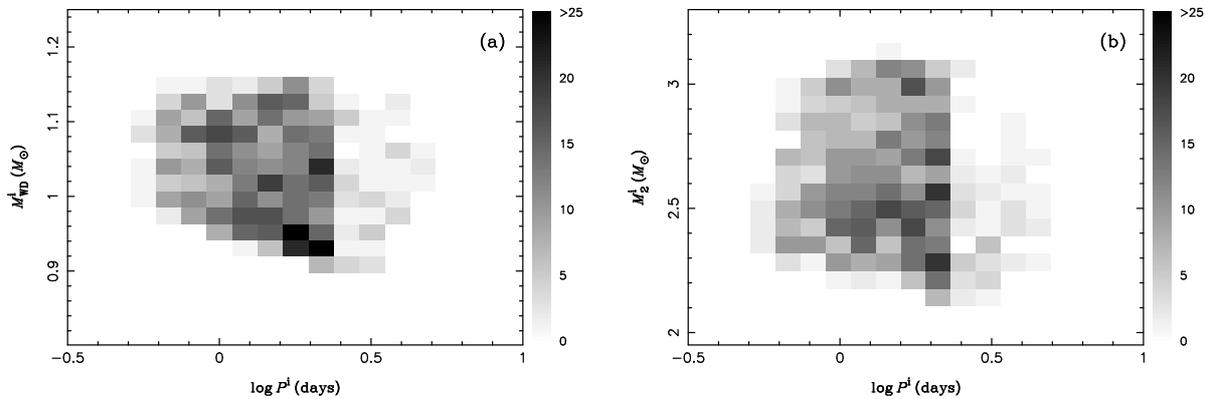

\centerline{\epsfig{file=f9a.ps,angle=270,width=8cm}\ \
\epsfig{file=f9b.ps,angle=270,width=8cm}} \caption{Distribution of properties of initial WD+MS systems
that can ultimately produce SNe Ia. Panel (a) shows the results in the $\log P^{\rm i}-M^{\rm i}_{\rm WD}$ plane, while panel (b) gives the results in the $\log P^{\rm i}-M^{\rm i}_{\rm 2}$ plane, where $P^{\rm i}$, $M^{\rm i}_{\rm WD}$ and  $M^{\rm i}_{\rm 2}$ are the initial orbital period, the initial WD masses, and the initial mass of the MS star, respectively.  Here, we present the case of set 1.}
\end{figure*}

In the observations, some WD+MS systems are suggested to be candidates of SN Ia progenitors
(for recent reviews see Wang \& Han 2012; Parthasarathy et al.\ 2007 and references therein).
In Fig.\,9, we give some properties of initial WD+MS systems that can
produce SNe Ia based on our BPS calculations, which would be useful for identifying potential progenitor candidates of
SNe Ia. For these WD+MS systems, SN Ia explosions in our simulations
occur for the ranges $M_{\rm WD}^{\rm i}\sim0.9$$-$$1.2\,M_\odot$,
$M_{\rm 2}^{\rm i}\sim2.1$$-$$3.1\,M_\odot$ and $P^{\rm i} \sim 0.5$$-$5\,days.
However, recent observations in the SDSS have not found massive MS stars in WD+MS systems, where
the MS stars have their masses  $<$$0.6\,M_\odot$
(e.g., Zorotovic et al. 2011; Nebot G\'{o}mez-Mor\'{a}n et al. 2011;  Rebassa-Mansergas et al.
2012, 2013; Toonen \& Nelemans 2013; Camacho et al. 2014; Li et al. 2014),
which cannot afford enough material for the WDs to increase their masses to the Chandrasekhar limit.

\section{Discussion and conclusions}

Previous studies usually assumed that, when the mass-transfer rate is lower than
$\frac{1}{2}\dot{M}_{\rm cr}$ but higher than $\frac{1}{8}\dot{M}_{\rm cr}$
(see dashed lines in Fig. 1),
a weak H-shell flash is triggered but no mass is lost from the binary system,
i.e., the H is totally converted into He during the mass-accretion phase
(e.g., Hachisu et al. 1999; Han \& Podsiadlowski 2004;
Meng et al. 2009; Chen \& Li 2009; Wang et al. 2010; Tauris et al. 2013).
However, in this paper we adopted a detailed mass-accumulation efficiency
for H-shell flashes once the mass-transfer rate is below $\dot{M}_{\rm st}$
(see Fig. 2), the value of which is lower than previous studies, leading to
a low birthrate of SNe Ia. Note that there is a heated debate about the mass-accumulation
efficiency on WDs; it is still uncertain whether nova outbursts remove
more mass from the WD than the accreted material (e.g., Prialnik \& Kovetz 1995;
Cassisi et al. 1998; Yaron et al. 2005; Idan et al. 2012; Newsham et al. 2013).

Denissenkov et al. (2013b) recently proposed a new type of WDs, i.e.,
hybrid CONe WDs, which have
an unburned CO core surrounded by a thick ONe shell. In a subsequent investigation,
Chen et al. (2014) found that this type of hybrid WDs could be as large as $1.3\,M_{\odot}$.
According to the works of Chen et al. (2014), Meng \& Podsiadlowski (2014) studied
the hybrid WD+MS channel of SN Ia progenitors based on a detailed BPS approach, and gave the birthrates
$\sim$$0.026-0.24\times 10^{-3}\,{\rm yr}^{-1}$. If this kind of hybrid WDs can really undergo a
thermonuclear runaway when increasing their masses to a critical limit,
then the SN Ia birthrates from the WD+MS channel can be increased significantly. Note
that hybrid WDs could reach a state of explosive
carbon ignition, but this depends on some mixing assumptions and the
convective Urca process  (see Denissenkov et al. 2015).

When the WD in the WD+MS channel explodes as an SN Ia, the mass donor star in this channel
would survive and potentially be identifiable (e.g., Liu et al. 2012b; Pan et al. 2014).
Tycho G was referred as the surviving companion of Tycho's SN from the
WD+MS channel (e.g., Ruiz-Lapuente et al. 2004; Wang \& Han 2010), although it is still debatable (e.g., Ihara et al.
2007; Kerzendorf et al. 2009; Liu et al. 2013).
The surviving companions from this channel may also provide a way to explain the formation of
single low-mass He WDs in the observations  (e.g., Justham et al. 2009; Wang \& Han 2010).

The process of mass accretion on CO WDs is important for understanding the SD model of SNe
Ia, which has a direct effect on the theoretical birthrates of SNe Ia (see also Bours et al. 2013).
In this article, by employing the Eggleton stellar evolution code and adopting
detailed mass-accumulation efficiencies of H-shell flashes on the WDs, we performed
detailed binary evolution calculations for the WD+MS channel based on the
super-Eddington wind scenario, and gave the initial parameter spaces for producing
SNe Ia with various initial WD masses. We found that a CO WD with its mass as low
as 0.9\,$M_{\odot}$ can potentially accrete enough mass and then
reach the Chandrasekhar mass limit if we adopt $\dot{M}_{\rm Edd}$ as the critical accretion rate. By
incorporating the results of the detailed binary evolution calculations into a detailed BPS approach, we
obtained the evolution of SN Ia birthrates with time. The SN Ia birthrate from the
WD+MS channel is much lower than the value obtained from observations.
This means that the contributions of the WD+MS channel to the total SN Ia
birthrates are quite inefficient. However, this result is strongly sensitive to
uncertainties in some initial conditions, especially the efficiency of mass accretion.
If we adopt $2\dot{M}_{\rm Edd}$ as the critical accretion rate,
the SN Ia birthrate is increased significantly. Additionally, the values of $\alpha_{\rm CE}$
also have an influence on the birthrates of SNe Ia. To set further constraints on the
WD+MS channel, large samples of observed massive WD+MS systems and surviving companions
are needed, and the process of mass accretion on WDs should be examined carefully.

\begin{acknowledgements}
We acknowledge the anonymous referee for valuable comments that helped us to improve the paper.
We thank Xiangdong Li, Fenghui Zhang and Xiangcun Meng for their useful discussions.
We also thank Yan Gao for his kind help to improve the language of this paper.
This study is supported by
the National Basic Research Program of China (973 program, 2014CB845700),
the National Natural Science Foundation of China (Nos 11322327, 11390374 and U1331117),
the Chinese Academy of Sciences (No. XDB09010202),
and the Natural Science Foundation of Yunnan Province (Nos 2013FB083 and 2013HB097).

\end{acknowledgements}

\clearpage

\end{document}